# Mass-manufactured Gradient Plasmonic Metasurfaces for Enhanced Mid-IR Spectrochemical Analysis of Complex Biofluids

Samir Rosas[1,‡], Shovasis Kumar Biswas[2,‡], Wihan Adi[1], Furkan Kuruoglu[1,3], Aidana Beisenova[1], Manish S. Patankar[4], Filiz Yesilkoy[1]*

[1] Department of Biomedical Engineering, University of Wisconsin–Madison, Madison, WI 53706, USA

[2] Department of Electrical and Computer Engineering, University of Wisconsin-Madison Madison, WI 53706, USA

[3] Department of Physics, Faculty of Science, Istanbul University, Vezneciler, 34134, Istanbul, Turkey

[4] Department of Obstetrics and Gynecology, University of Wisconsin–Madison, Madison, WI 53792, USA

*Corresponding author. Email: filiz.yesilkoy@wisc.edu

‡ Authors contributed equally

## Abstract

Mid-infrared spectroscopy offers powerful label-free molecular analysis capabilities but faces significant challenges when analyzing complex biological samples. Here, we present a transformative surface-enhanced infrared absorption spectroscopy (SEIRAS) platform that overcomes fundamental limitations through key innovations. First, we demonstrate high-throughput wafer-scale fabrication of mid-IR plasmonic micro-hole-array (MHA) metasurfaces on free-standing silicon nitride membranes, yielding approximately 400 sensor chips per 6-inch wafer. Second, our gradient MHA metasurface design supports spectrally cascaded plasmonic modes, generating over 400 sharp resonance peaks across the 1200-2000 cm$^{-1}$ fingerprint region. This approach enables comprehensive molecular fingerprinting using simple imaging optics in transmission mode. Third, we validate our SEIRAS platform using a model polymer system and clinical peritoneal fluid samples from ovarian cancer patients, demonstrating its capability to resolve complex molecular signatures in real biological specimens. The platform's dense spectral coverage ensures optimal on-resonance enhancement across the broad fingerprint region, revealing previously obscured vibrational bands that conventional IR spectroscopy cannot distinguish. By combining high-throughput fabrication with simplified optical readout and the capability to analyze complex biological samples, our work establishes a foundation for translating SEIRAS technology into practical biomedical applications.

# INTRODUCTION

Label-free vibrational spectroscopy in the mid-infrared (mid-IR) spectrum ($\lambda$ = 2.5 – 25 µm) is a powerful tool for noninvasive analysis of biological samples without lengthy sample staining and preparation. This technique enables the detection of biomolecules based on their unique spectral fingerprints, where molecular vibrational energy states are inscribed as absorption peaks[1,2]. Spectrochemical fingerprinting is particularly promising for biomedical diagnostics[3,4], disease monitoring[5,6], and biomarker discovery[7,8] because it can simultaneously capture diverse molecular structures from biospecimens. However, when analyzing complex biological samples, conventional mid-IR absorption spectroscopy (MIRAS) faces significant challenges. Specifically, heterogeneity in sample thickness and composition[9] and optical scattering[10] distort and congest spectral data, impeding reliable chemometric analyses. Therefore, innovative analytical platforms capable of capturing accurate mid-IR spectral fingerprints from real-world samples are urgently needed.

To overcome these limitations, surface-enhanced infrared absorption spectroscopy (SEIRAS) has been developed, offering significantly higher sensitivity through enhanced light-matter interactions in nanophotonic cavities[11,12]. To date, numerous SEIRAS technologies have been proposed using antennas[13,14], antenna clusters[9,15], and their 2D arrays forming metasurfaces from metallic[16], dielectric[17–19], and 2D materials[20]. By leveraging strong near-field light localization, SEIRAS has been applied to analyze biological cells[21], tissues[9], and functional biomolecules, such as proteins[22], nucleic acids[23], carbohydrates[24]. However, current SEIRAS platforms face three major challenges that limit their impact and hinder their practical applications.

First, nanostructured SEIRAS substrates typically rely on low-throughput and expensive electron-beam lithography fabrication on IR-transparent windows like $CaF_2$ and $MgF_2$. Since these substrate materials are brittle and incompatible with wafer-scale techniques, their technology transfer into mass-produced products has been hindered. Second, SEIRAS signals are usually embedded in the far-field resonance distortions, such as amplitude damping, frequency shifting, and mode broadening. Typical mid-IR antennas generate resonance peaks in the reflection mode requiring complex readout platforms, which obstruct the development of cost-effective, robust, and portable sensors. Third, when measuring chemically complex samples using SEIRAS, accurate retrieval of absorption spectra in the broad fingerprint region requires alignment of photonic resonances to individual vibrational bands. This alignment is essential because the electromagnetic near-field enhancement reaches its maximum at the central wavelength of the resonance peak. Furthermore, achieving on-resonance SEIRAS ensures realistic symmetric band retrieval, in contrast to Fano-like asymmetric profiles that emerge when the photonic resonance is detuned from the vibrational band[25].

Some of these SEIRAS challenges have been addressed by recent developments in the nanophotonics field. For example, to eliminate the reliance on IR-transparent substrates, Si[17,18] and $Al_2O_3$ membranes[26] were used to fabricate metasurfaces for SEIRAS applications. To enable simple SEIRAS sensor platforms operating in transmission mode, we recently demonstrated transmissive resonances supported by Si membrane[17] metasurfaces. Moreover, others proposed plasmonic devices with transmission resonances for SEIRAS applications[27–29]. To enable on-resonance SEIRAS across the fingerprint spectrum, gradient metasurfaces were developed where continuously tuned metasurface resonances enabled precise alignment with multiple vibrational bands on the same chip[30,31]. While these previous efforts tried to address individual limitations of the SEIRAS systems, a single platform to overcome all the problems at once has not yet been reported. Additionally, previous platforms are often validated using pure model molecules, such as PMMA or protein and DNA species, rather than complex biological samples, leaving their real-world utility unproven.

Here, we present a transformative SEIRAS platform that circumvents the fundamental limitations through key innovations in a single device. First, we demonstrate wafer-scale, high-throughput fabrication of plasmonic micro-hole-array (MHA) metasurfaces on free-standing silicon nitride ($Si_3N_4$) membranes, yielding approximately 400 sensor chips per 6-inch wafer (Fig. 1a). Second, our SEIRAS platform enables transmission mode optical readout because the MHA resonance mechanism is driven by extraordinary optical transmission (EOT), where the interaction between dark surface plasmon polaritons and bright localized surface plasmon modes results in a spectral peak in transmission[32]. Third, our gradient MHA design generates a dense spectrum of resonance modes (>400 distinct modes), enabling on-resonance fingerprint retrieval across the 1200-2000 $cm^{-1}$ fingerprint region. We validated our platform using a model polymer system (PMMA), highlighting the critical impact of resonance detuning on SEIRAS-retrieved absorbance band profiles. Moreover, we analyzed human peritoneal fluid samples from ovarian cancer patients, demonstrating the real-world utility of our SEIRAS approach in solving complex molecular signatures in biological specimens. Overall, this work introduces a practical solution for implementing SEIRAS in real-world applications, offering a scalable platform for detailed molecular analysis of biological samples. The combination of high-throughput fabrication, comprehensive spectral coverage, and a simple transmission-mode optical readout represents a significant advance toward making SEIRAS a viable tool for medical diagnostics and biomarker discovery.

## RESULTS

### High-throughput fabrication of gradient plasmonic micro-hole-array (MHA) metasurface chips

The high-throughput wafer-scale fabrication of plasmonic MHA gradient metasurfaces begins on a 6-inch (150 mm) silicon wafer. First, a 400 nm silicon nitride ($Si_3N_4$) layer is deposited using low-pressure chemical vapor deposition (LPCVD), followed by the application of an antireflective coating (ARC) and a UV-sensitive positive photoresist. Next, the gradient MHA pattern is generated by standard photolithography (using a 365 nm wavelength), and the holes were etched into the $Si_3N_4$ layer using reactive ion etching (RIE) (Fig. 1a). To release the patterned $Si_3N_4$ membranes, backside of the wafer was lithographically patterned, and the Si was etched away from the membrane areas. Each MHA chip features three suspended $Si_3N_4$ windows, each measuring a 3.0 mm × 0.7 mm area (Fig. 1b). Approximately 400 sensor chips are retrieved from each fabricated wafer. Further fabrication details are included in the methods section and in[33]. Once the fabrication of the wafer is complete, individual chips measuring 5.4 mm × 5.4 mm are separated and coated with a 100 nm-thin gold (Au) layer to induce the plasmonic effect (Fig. 1c). A scanning electron microscope (SEM) image of the fabricated MHA with a hexagonal lattice is shown in Fig. 1d. Figure 1e presents a photograph of the gradient MHA metasurface, which consists of 30 hexagonal hole-array patterns with varying geometric parameters ($P_x$, $P_y$, r), labeled $MS_1$ through $MS_{30}$ on a single chip. The side view of the plasmonic MHA gradient metasurface (Fig. 1f) shows a silicon frame (310 μm thick) for easy handling and an MHA-patterned $Si_3N_4$ layer (400 nm) providing robust mechanical support for the top gold layer. Upon illumination with a broadband mid-IR light source, each of the 30 metasurfaces exhibits a resonance peak in transmission at a specific wavenumber, covering the range of 1200 cm$^{-1}$ to 2000 cm$^{-1}$, as shown in Fig. 1g.

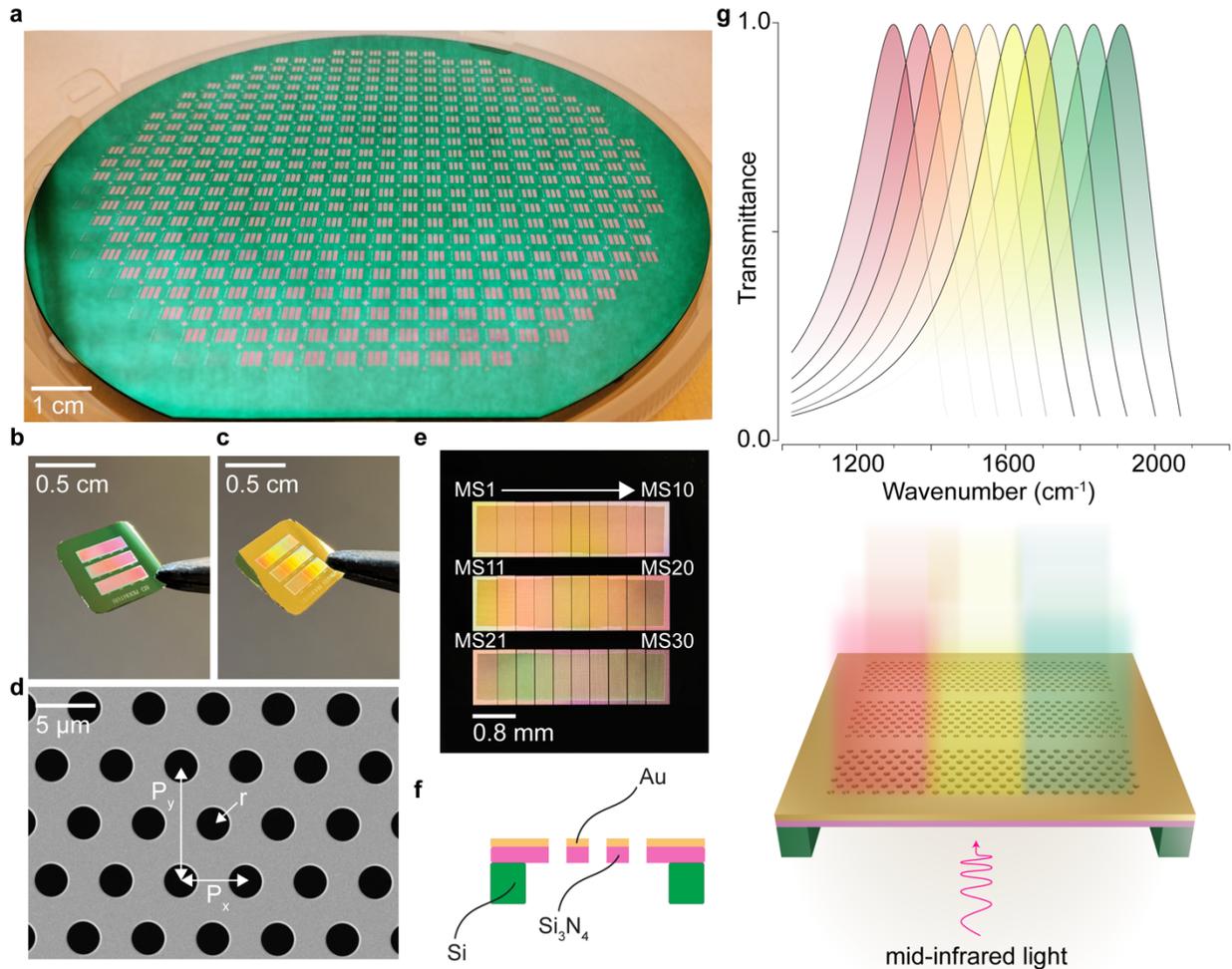

**Figure 1. High-throughput-manufactured gradient micro-hole array (MHA) metasurfaces. a,** Photograph of a 6-inch wafer carrying ~400 chips, each patterned with three silicon nitride ($Si_3N_4$) membranes (3 mm × 0.7 mm). Hole arrays with gradually changing dimensions are lithographically fabricated on each membrane. Photographs of individual chips before (**b**) and after Au deposition (**c**). **d,** Scanning electron microscopy (SEM) image of an Au-coated hexagonal lattice MHA with geometrical parameters $P_x$ = 5.94

µm, $P_y$ = 10.70 µm, and r = 1.57 µm. **e**, Photograph of a gradient plasmonic MHA metasurface chip, which includes 30 different metasurfaces ($MS_1$..., $MS_{30}$) with varying MHA geometric parameters. **f**, A cartoon cross-sectional view of the chip shows the material stack with a top Au layer (100 nm) deposited on a $Si_3N_4$ free-standing membrane (400 nm), which is supported by a 310 µm-thick silicon substrate. **g**, Au-coated, gradient MHA patterned, three parallel $Si_3N_4$ membranes on each sensor chip support the excitation of a comb of plasmonic extraordinary transmission resonances. Each metasurface exhibits a unique plasmonic resonance peak in transmission, uniformly covering a spectral range of 1200 $cm^{-1}$ to 2000 $cm^{-1}$ in the mid-infrared fingerprint spectrum.

## Plasmonic resonance properties of micro-hole array (MHA) metasurfaces

The plasmonic MHA metasurfaces exhibit the distinctive extraordinary optical transmission (EOT) phenomenon[32] driven by the interplay of two resonance mechanisms that produce an asymmetric Fano-type spectral line profile in transmission. When the MHAs are illuminated normally, in-plane surface plasmon polariton (SPP) modes are excited, satisfying the Bragg's coupling condition through wavevector matching with the grating's momentum. This SPP dark mode interacts strongly with the subwavelength holes in the Au film, generating localized surface plasmon (LSP) modes that scatter light into free space. The near-field interactions between the dark SPP and bright LSP modes are highly dispersive, giving rise to the Fano-type transmission peak in the far-field intensity spectrum (Fig. 2).

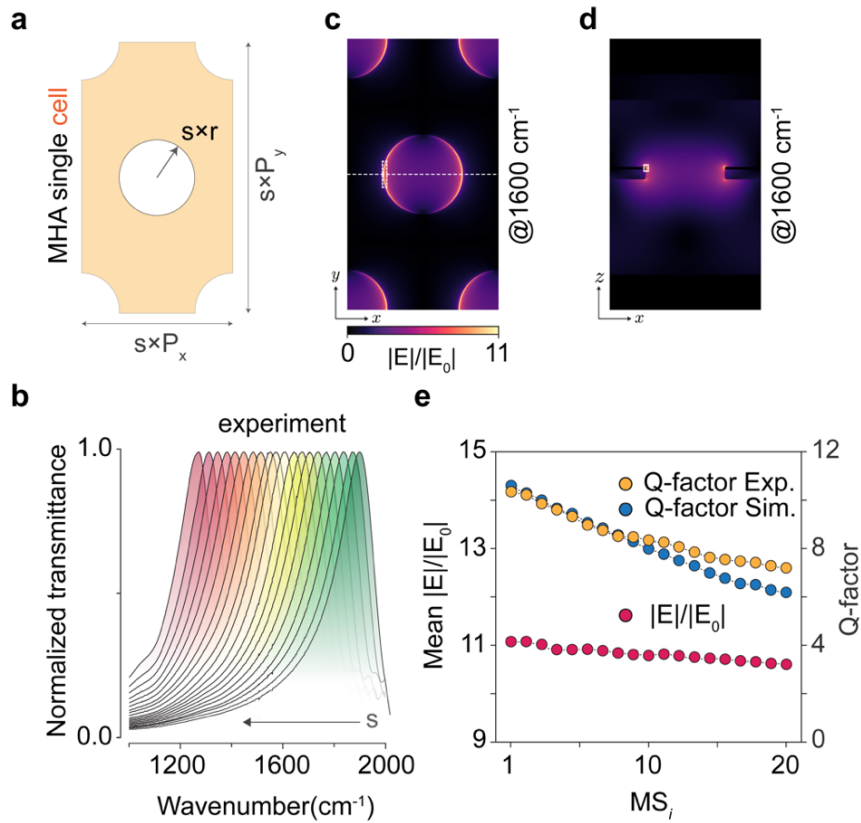

**Figure 2. Photonic resonance characteristics of the gradient MHA metasurface**. **a**, Schematic of an Au-MHA unit cell used in simulation to generate a hexagonal lattice with base periods $P_x$ = 5.94 µm and $P_y$ = 10.70 µm, and a hole radius r = 1.57 µm. These parameters are scaled to sweep the resonance across the desired spectral range. **b**, Measured transmission spectra of a gradient Au-MHA metasurface. By adjusting the scaling factor S in (**a**), we tune the wavenumber of the resonance peak from 1200 cm$^{-1}$ (S = 1.38) to 2000 cm$^{-1}$ (S = 0.8). Electric field enhancement, $|E|/|E_0|$, maps of the top (**c**) and side (**d**) views of a unit cell at the resonance wavenumber ($\lambda_{res} = 1600$ cm$^{-1}$) showing strong field localization around the rims of the holes. The calculated effective mode volume at the resonance peak is $V_{eff} = 0.0022(\lambda/n_{eff})^3$ with $n_{eff} = 3.033$. **e,** Metasurface resonance frequency dependent Q-factor variation calculated from experimentally measured (yellow) and simulation-derived (blue) resonance spectra. The metasurface-dependent E-field enhancement is determined by averaging $|E|/|E_0|$ over the volume indicated by the dashed rectangular regions shown in (**c**) and (**d**).

The EOT resonances can be spectrally tuned by modulating the unit cell geometrical parameters, i.e., period along the x-axis ($P_x$ = 5.94 µm), period along the y-axis ($P_y$ = 10.70 µm), as well as the hole radius (r = 1.57 µm), using a scaling factor (S) (Fig. 2a). In our gradient MHA metasurface, these EOT resonances sweep the mid-IR fingerprint region with a dense comb of transmission peaks (Fig. 2b). This is achieved by varying the S-factor from 1.38 to 0.8 over a length of 9 mm to cover a spectral range of 1200–2000 cm$^{-1}$. Overall, the measured transmittance response is in excellent agreement with the simulation (see SI, Fig. S1). Moreover, from the electric field maps (Fig. 2c top face, and Fig. 2d, transversal view), we see that the LSP mode is tightly bound to the rims of the holes, extending into the accessible hole openings (Fig. 2d) with a maximum electric field enhancement of $|E|/|E_0|$~11 at 1600 cm$^{-1}$. Fig. 2e shows the resonance quality factor (Q-factor) and the electric field enhancement (mean $|E|/|E_0|$ from a volume of $16 \times 10^{-3}$ µm$^3$) variations in the gradient MHA metasurfaces. When comparing the simulated and experimental responses of the MHA gradient, a decrease in the Q-factor (~36% across 20 MS) is observed with increasing resonance wavelength (increasing S-parameter). Similarly, the E-field enhancement decreases slightly (~10% across 20 MS), enabling stable sensing across the fingerprint spectrum.

We further investigated the spatial variation of the resonance properties in the gradient MHA metasurfaces using a mid-IR hyperspectral imaging microscope. Coupled with a tunable quantum cascade laser (QCL), we collected hyperspectral image data cubes by sweeping collimated, normally incident, and linearly polarized mid-IR light between 5.5 µm and 10.5 µm. From each metasurface area, we considered 200×200 pixel-spectra and plotted their resonance wavelengths on a density plot (Fig. 3). We measured minimum and maximum standard deviations of 7 cm$^{-1}$ and 13 cm$^{-1}$, respectively, for gradients MS$_{5-13}$. From the density plot, we observed that our fabricated gradient MHA metasurfaces densely populate the spectral region of interest, with at least 20 wavenumbers of overlap in pixel-resonance distribution between adjacent metasurfaces. This overlap is especially advantageous for retrieving vibrational bands with minimal distortion, as will be discussed in the next sections. Additionally, Fig. 3 shows three infrared transmission microscopy images of the MHA gradient chip, revealing three parallel patterned membrane regions. These distinct regions light up sequentially from top to bottom, in accordance with their resonance position, as the illumination wavenumber changes (1200, 1500, and 1800 cm$^{-1}$).

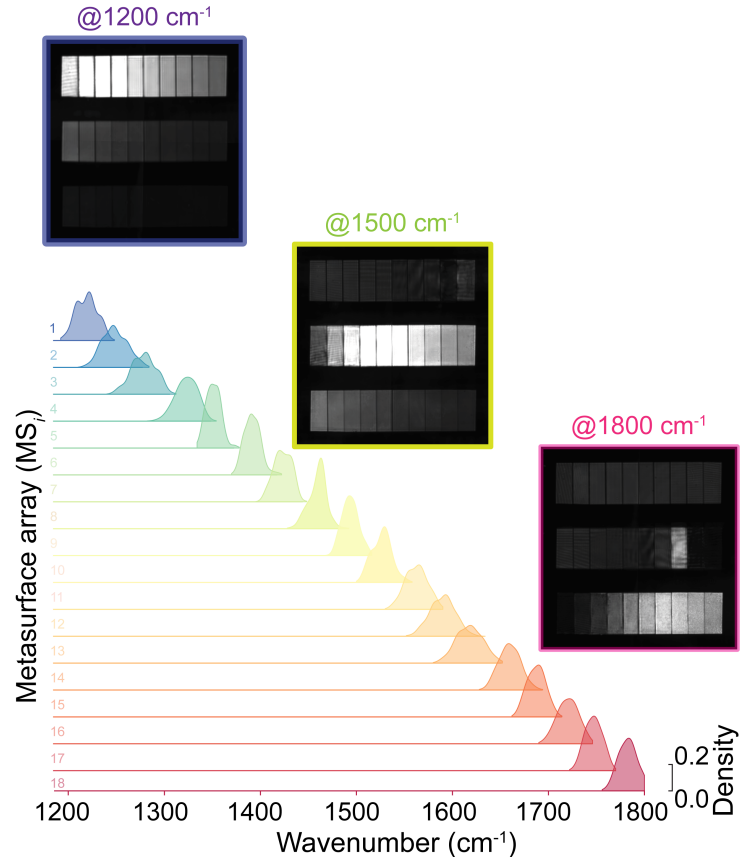

**Figure 3. Spatial distribution of resonance peak positions in the gradient Au-MHA metasurface.** Hyperspectral imaging-based optical interrogation was employed to analyze the density functions of spatial resonance peak distributions across 18 metasurfaces (each sampled by 200 × 200 pixels). The insets show three infrared transmission microscopy images acquired at 1200 cm$^{-1}$ (top), 1500 cm$^{-1}$ (middle), and 1800 cm$^{-1}$ (bottom). Each rectangular metasurface pattern measures 3.0 mm × 0.7 mm.

### *On-resonance SEIRAS using plasmonic gradient MHA metasurfaces*

To demonstrate the critical role of on-resonance SEIRAS for chemical characterization of complex compounds, we investigated the effects of spectral proximity between the plasmonic resonance and the carbonyl vibrational band (C=O at 1730 cm$^{-1}$) of PMMA molecules affects signal quality. We began by measuring the spectral response of a bare MHA gradient metasurface to establish a baseline. Subsequently, a 70 nm-thick PMMA layer was spin-coated onto the top face of the same gradient MHA chip, as artistically rendered in Fig. 4a. Figure 4b shows the measured absorbance spectra of the coupled system, comprising the EOT plasmonic resonance and the carbonyl band of PMMA, from multiple MHA metasurfaces. To extract the metasurface-enhanced absorbance spectra of the carbonyl band, we subtracted the transmittance baseline of the bare MHA from that of the MHA-PMMA coupled system (Fig. 4b insets). When the MHA resonance peak is finely aligned with the carbonyl vibrational mode (Fig. 4b ii), the retrieved absorption band displays an almost symmetrical shape and fits well to a Gaussian curve. In contrast, when the resonance peak is detuned either to the left or right of 1730 cm$^{-1}$, the retrieved carbonyl band becomes asymmetric, exhibiting a Fano-like profile (Fig. 4b i, iii, and iv). To quantify detuning effects, we calculated the degree of asymmetry using a Fano fit (details in Methods) and plotted the asymmetry factor ($\alpha$) as a function of detuning (Fig. 4d). We also calculated the maximum absorbance as a function of spectral detuning (Fig. 4c), and, as expected, found that the maximum enhancement occurs at zero detuning point, where the coupling is highest. This experiment demonstrates not only the impact of the resonance peak alignment on accurately and sensitively retrieval of the vibrational bands but also highlights the importance of our SEIRAS approach using gradient metasurfaces for precise chemical analysis of complex samples.

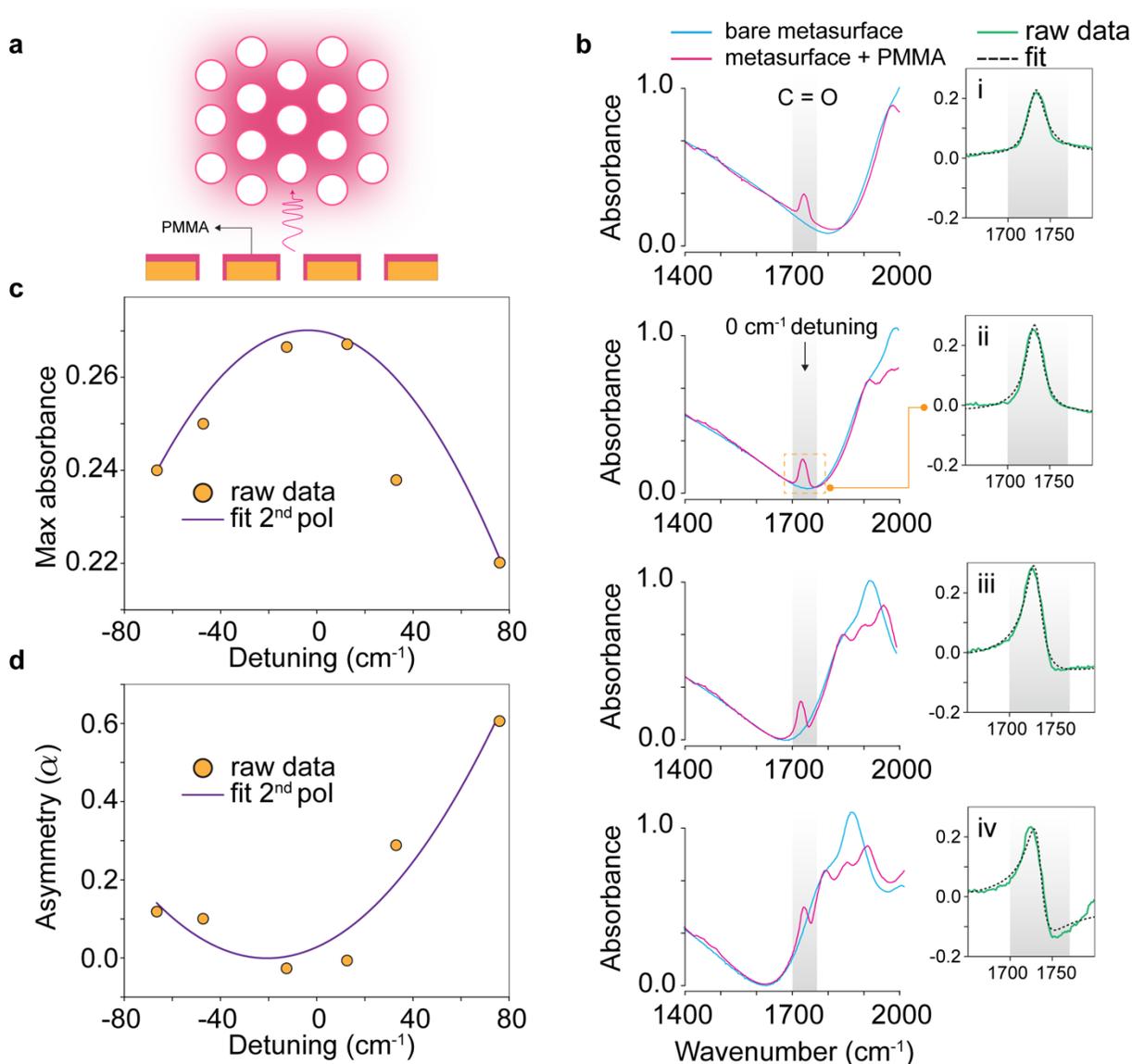

**Figure 4. On-resonance surface-enhanced Infrared absorption spectroscopy (SEIRAS) with gradient plasmonic Au-MHA metasurfaces. a**, A 70 nm-thick PMMA layer was spin-coated on the top face of the gradient Au-MHA metasurface. **b**, Measured absorbance spectra of metasurfaces with varying resonances before (blue) and after (pink) PMMA coating. The insets show the retrieved PMMA absorbance band obtained by subtracting the bare metasurface spectra from those of the PMMA-coated metasurface. As the bare resonance of the metasurface detunes from the 1730 cm$^{-1}$ C=O vibrational band of PMMA molecules, the absorbance profile deviates from Gaussian to Fano-like shape. **c, d,** Calculated maximum

absorbance (**c**) and Fano factors (**d**) from the SEIRAS retrieved C=O absorbance bands in (**b**) indicate that, at zero detuning(**b**-ii), the C=O absorbance band is highly symmetric, highlighting the importance of gradient metasurfaces in SEIRAS for accurately detecting realistic absorption band profiles.

An important application field for vibrational spectroscopy is biomedicine, where non-targeted, label-free chemical analysis of complex biological samples is required. However, in real-world samples, accurate identification of spectral fingerprints in the mid-IR region suffers from spectral congestion and non-specific variations due to measured samples' physical features[9]. As an example, we analyzed a human peritoneal fluid (PF) sample from an ovarian cancer patient (see Methods) by drop casting a 5 µL of PF on a standard $CaF_2$ substrate. Figure 5a shows a mid-IR absorption image captured at the Amide I band (1656 cm$^{-1}$) using a standard transmission-mode measurement. Herein, we demonstrate the nonuniform PF sample distribution forming a coffee-ring stain as it dries on the $CaF_2$ surface. In Fig. 5a, we further plot the absorbance spectra from four different spatial positions (1,2,3, and 4) of the same sample. Spectra from positions 1 and 2 do not saturate because these regions are thinner, yet the weak bands are not accurately captured. By contrast, the spectra from thicker regions saturate the strong amide bands, and the Mie scattering generates a large baseline[10]. Overall, this experiment demonstrates that the spectral fingerprints exhibit distinct quantitative and qualitative differences, despite being measured from the same PF sample, highlighting the challenges of accurate spectral measurements of real-world samples.

For accurate chemical analysis of complex biological samples, we developed gradient metasurfaces whose plasmonic resonances continuously sweep the mid-IR spectral range, enabling on-resonance SEIRAS across the fingerprint region. Our method begins with spin-coating a 50 µL of PF at 4000 rpm onto the top surface of the Au-MHA gradient metasurface (see Methods). Figure 5b presents a mid-IR absorption image of thinly coated PF on the Au-MHA metasurface, recorded in transmission mode at the Amide I band (1656 cm$^{-1}$). We further processed the spectra from 200×200 number of pixels in this region, plotting their mean and standard deviation (Fig. 5b). The absorption bands captured across the metasurface remain quantitatively and qualitatively uniform, highlighting the strength of SEIRAS in characterizing thinly spread complex biological samples.

Using the same rationale as in our coupled MHA-PMMA detuning study, we investigated how the detuning affects the SEIRAS response of a complex human PF sample. Specifically, we compared spectra from four adjacent metasurfaces ($MS_{18}$, $MS_{17}$, $MS_{16}$, and $MS_{15}$) whose resonances sweep through the protein-associated amide I and II spectral regions (gray shaded areas in Fig. 6). Figure 6a shows mid-IR transmittance images of the four metasurfaces spin-coated with PF, captured at the amide I band (1656 cm$^{-1}$). We also plotted the average transmittance spectra from these images in Fig. 6b. Vibrational bands associated with functional groups in biomolecules distort the plasmonic EOT resonances, suppressing transmittance peak in the far-field. We observed that these vibrational bands appear asymmetric when they are not spectrally aligned with the resonance peak of the plasmonic MHA metasurfaces, similar to the behavior seen in the MHA-PMMA coupled system. Moreover, because the PF exhibit more complex bands than those of the PMMA molecules, we analyzed the PF fingerprint spectra by taking the second derivative of its absorbance signals from the coupled MHA-PF system. Figure 6c compares these second-derivative spectra from four adjacent MHA metasurfaces with the reference spectrum of the same PF sample collected using a standard transmission-mode measurement.

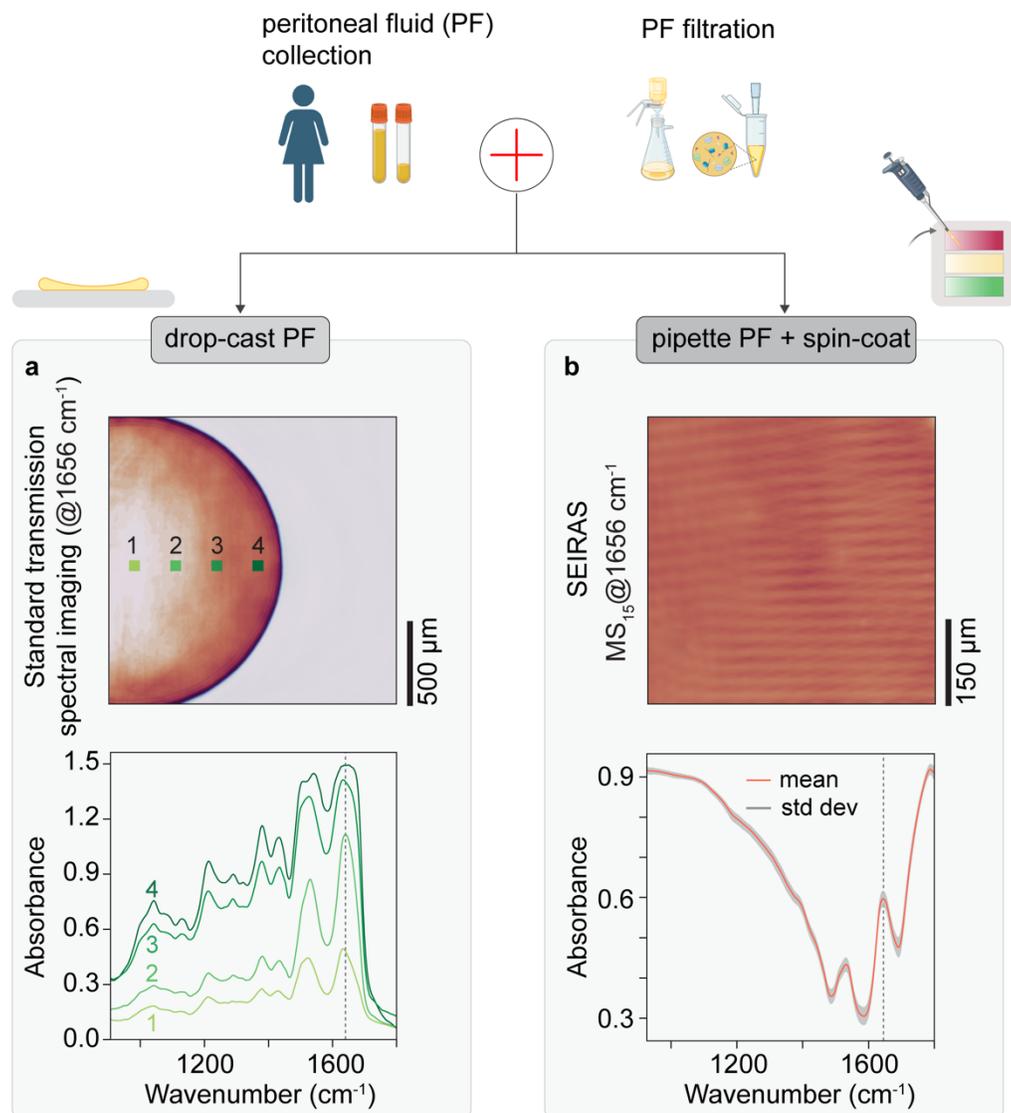

**Figure 5. Mid-IR spectral analysis of human bodily fluids. a**, Mid-IR transmission microscopy image of drop-cast, dried peritoneal fluid on a standard CaF$_2$ substrate, captured at the Amide I protein band (1656 cm$^{-1}$) using a hyperspectral imaging instrument. The absorption image reveals uneven PF sample thickness due to the coffee-ring effect. Spectra from four different spots along the radial axis of the droplet (1,2,3, and 4) exhibit drastic quantitative and qualitative variations in the absorbance fingerprints, despite originating from the same sample. **b**, Mid-IR transmission microscopy image of the spin-coated PF sample on a metasurface, which exhibits uniform absorbance unlike the drop-casted PF. In the SEIRAS spectra, the red line represents the mean value of the spectra collected from a metasurface area of 90×10$^3$ μm$^2$, while the shaded gray region corresponds to the standard deviation. Despite the thin PF layer, the plasmonic metasurface enhances the absorbance signals, enabling the acquisition of consistent fingerprint spectra of the sample.

Our second-derivative analysis revealed that when molecular absorption bands are probed with on-resonance SEIRAS, their amplitudes and spectral positions are better resolved than in the detuned cases. For example, vibrational bands around 1518 cm$^{-1}$, 1546 cm$^{-1}$, 1590 cm$^{-1}$, and 1644 cm$^{-1}$ are accurately retrieved by the MS$_{15}$ (Fig. 6c, magenta) and their spectral positions show good agreement with the reference signal collected from bulk PF sample on CaF$_2$ substrate (Fig. 6c, grey). In contrast, these bands appear weak in the spectra from the adjacent metasurfaces (MS$_{16}$, MS$_{17}$, MS$_{18}$) and their spectral positions vary due to the asymmetry introduced by resonance detuning. Likewise, the bands of 1660 cm$^{-1}$ and 1686 cm$^{-1}$ (Fig. 6c, green) are most accurately captured by MS$_{16}$. Moreover, our measurements unveil how SEIRAS addresses the issue of spectral congestion in standard mid-IR spectroscopy. For instance, the two closely spaced bands (1640 cm$^{-1}$ and 1656 cm$^{-1}$) are barely distinguishable in the grey reference spectra in Fig. 6c, yet they are well separated in measurements with MS$_{15}$ and MS$_{16}$. This occurs because, in bulk sample measurements, the strong absorption from abundant molecules overshadows weaker bands. Notably, SEIRAS minimizes spectral congestion because the measurement volume is confined to plasmonic hotspots, where a small quantity of molecules can generate a high signal-to-noise ratio absorption signal due to enhanced light-matter interactions.

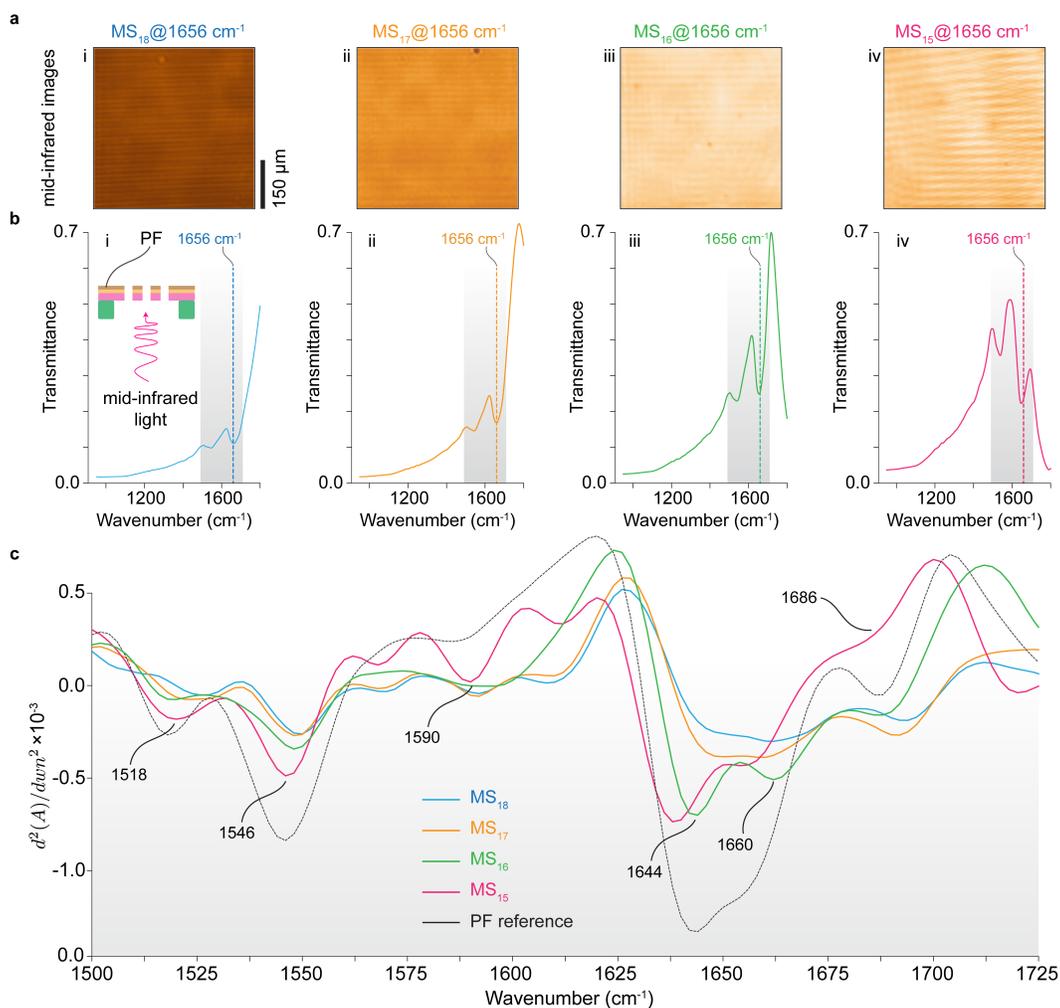

**Figure 6. Spectrochemical analysis of complex peritoneal fluids using the plasmonic gradient metasurfaces. a**, Mid-IR transmission microscopy images of the Au-MHA metasurfaces, MS$_{18}$, MS$_{17}$, MS$_{16}$,

and MS$_{15}$, spin-coated with a thin layer of PF on the top face. Changes in the transmittance intensity at 1656 cm$^{-1}$ are due to differences in the metasurface resonance positions. **b**, Measured transmittance spectra of PF-coated metasurfaces, whose resonances sweep the amide I and II regions (highlighted in gray). The absorbance enhancement becomes more prominent when the protein-associated amide bands are tuned to the resonance peaks. **c**, Second derivative analysis shows that sub-vibrational bands are better resolved when they are aligned with the metasurface resonance, exhibiting minimal distortion in their band profiles compared to the PF reference signal.

**DISCUSSION**

In this work, we present several key advances in SEIRAS technology for analyzing complex biological samples. By fabricating plasmonic gradient metasurfaces on free-standing Si$_3$N$_4$ membranes using wafer-scale processes, we achieved an unprecedented manufacturing throughput (~400 chips per 6" wafer). This accomplishment critically overcomes the low-throughput challenge in chip fabrication and makes SEIRAS technology practically viable for widespread adoption in real-world sensing applications.

Second, our transmission-mode operation, enabled by extraordinary optical transmission of the plasmonic MHA metasurfaces, simplifies the optical readout setup. Our sensor chips allows optical interrogation using robust, collinear imaging platforms consisting of a broadband IR light source, a microbolometer IR camera, and simple lenses, as previously demonstrated by[34]. Such straightforward and compact analytical platforms can further broaden the impact of SEIRAS by facilitating its implementations as on-site sensors for real-time monitoring.

Third, the gradient metasurface design, hosting over 400 distinct resonance modes across the fingerprint region, ensures optimal enhancement of molecular absorption signatures. As highlighted by our detuning studies with PMMA and PF, precise alignment of plasmonic resonances to the molecular absorption bands is crucial for accurate spectral retrieval. Thus, our gradient metasurface design ensures proper resonance overlap across the entire fingerprint region, which is particularly important when analyzing complex biological samples containing multiple overlapping bands.

Finally, we demonstrated our technology's real-world utility by successfully analyzing peritoneal samples from ovarian cancer patients. Unlike conventional IR spectroscopy, in which physical properties of the sample and scattering effects can distort spectral features, our SEIRAS approach provided consistent spectral measurements, underscoring its potential for biochemical characterization of clinical samples.

In sum, we addressed the key limitations of traditional SEIRAS methods through innovative design and fabrication approaches, representing a significant step toward practical, high-throughput and compact biochemical sensors. The combination of scalable fabrication of plasmonic chips, simplified imaging-based optical readout in transmission mode, and the capability to characterize complex biological samples establishes a foundation for translating SEIRAS technology into routine medical diagnostic and biomedical research applications.

## MATERIALS AND METHODS

### Numerical simulations

The finite-element frequency-domain solver (CST Microwave Studio 2023, Dassault Systèmes, France) was used to calculate the transmittance spectra in Figure 1g, and 2e. The electric field enhancement, depicted in Figure 2c and 2d, was calculated using Tidy3D FDTD (Flexcompute, California, USA). In all simulations, periodic boundary conditions were applied along the x- and y-axes, while perfectly matched layers (PML) were used along the z-direction. We assumed plane wave illumination with a Gaussian temporal profile, incident normally from the top of the structure. The optical constants for Au and $Si_3N_4$ were taken from references[35,36], respectively. The cavity's effective mode volume was calculated as described in reference[37].

$$V_{eff} = \frac{\int \varepsilon(\boldsymbol{r})|\boldsymbol{E}(\boldsymbol{r})|^2 d^3\boldsymbol{r}}{\max\left(\varepsilon(\boldsymbol{r})|\boldsymbol{E}(\boldsymbol{r})|^2\right)}$$

### Device fabrication

The gradient MHA metasurfaces were fabricated on a 150 mm silicon wafer, 310 µm thick ⟨100⟩ p-type, doubled-side-polished silicon wafers. A 400 nm silicon nitride ($Si_3N_4$) layer was deposited using low-pressure chemical vapor deposition (LPCVD) onto both the top and bottom faces of the wafer. This was followed by the application of an antireflective coating (ARC) and a UV-positive photoresist onto the wafer's front side. After conventional photolithography (365 nm wavelength) patterning, the MHA gradient geometry was transferred into the ARC via $Ar/O_2$ plasma etching and subsequently into the $Si_3N_4$ layer using reactive ion etching with $SF_6$ and He. To serve as a protective layer, a 200 nm $SiO_2$ coating was deposited over the MHA-patterned $Si_3N_4$.

The backside of the wafer was then patterned with a mask to define the chip's outer frame dimensions (5.4 mm × 5.4 mm) and three free-standing suspended membrane windows (3.0 mm × 0.7 mm). Finally, the bulk silicon was removed using wet etching. Further details of the fabrication process can be found in reference[33].

### PMMA spin coating and thickness characterization

A 70 nm-thick PMMA layer (PMMA 950 A2, Kayaku Advanced Material, Massachusetts, USA) was deposited using spin coating at 4000 rpm. The metasurface chip was then baked at 180°C for 5 minutes. To measure the PMMA thickness, each metasurface coated with PMMA was accompanied by a bare Si chip of the same size, which was also coated with PMMA under identical conditions. The thickness of the PMMA layer on the bare Si substrates was measured using an ellipsometer (J.A. Woollam, Nebraska, USA) and a reflectometer (Filmetrics, California, USA).

### Peritoneal fluid coating

Human peritoneal fluid (PF) was collected from patients with high-grade serous ovarian cancer as part of their standard-of-care treatment. These studies were approved by the Institutional Review Board (IRB), and all patients provided informed consent prior to PF collection. The samples were stored at –80°C within two hours of collection. To minimize aggregates in suspension—which could rupture the membrane due to mechanical stress during drying—the freshly collected PF was filtered using a Corning 50 mL vacuum filter with a 0.22 µm pore size. The absorbance spectrum was measured both before and after filtration. No significant change was observed in the bulk absorbance spectrum profile between the filtered and non-filtered PF samples. Next, 50 µL of PF was pipetted onto the top surface of the MHA gradient metasurface and incubated for five minutes at room temperature before being spin-coated at 4000 rpm. The chip was then left to dry for five hours at room temperature before being placed in the nitrogen-purged chamber of the mid-infrared microscope for spectral measurements. For the drop-casted PF, 5 µL of filtered PF was deposited onto a standard mid-infrared substrate. It was allowed to dry for five hours at room temperature before mid-infrared measurements were performed.

**Optical setup and measurements**
Mid-infrared spectral measurements were performed using a tunable quantum cascade laser (QCL) integrated into a mid-infrared microscope (Spero-QT, Daylight Solutions, California, USA), unless otherwise noted. The microscope, equipped with four QCL modules, can collect spectra covering the fingerprint region from 950 to 1800 cm$^{-1}$ with a spectral resolution of 2 cm$^{-1}$. During acquisition, the sample chamber was continuously purged with dry nitrogen to remove water vapor. The free-standing metasurface was illuminated with collimated, linearly polarized light at normal incidence. Spectral data were acquired in transmission mode using a 12.5× IR collection objective (0.7 NA) and detected using an uncooled microbolometer focal plane array with 480 × 480 pixels, providing a field of view of 650 µm × 650 µm. In transmission mode, our chemical microscope captures the sample's decadic absorbance as:

$$A = -\log_{10}(T)$$

Where $T = I/I_\circ$ represents the transmission through a standard CaF$_2$ substrate. From the definition of decadic absorbance, the transmittance of the sample is calculated as:
$$T = 10^{-A}$$
Figure 2b (experiment) and Figure 4b were composed using data collected with a Fourier-transform infrared (FTIR) spectrometer coupled to an infrared microscope (Bruker Vertex 70 FTIR and Hyperion 2000). Transmittance spectra were obtained using linearly polarized light and a low-NA refractive objective (5×, 0.17 NA, Pike Technology, Wisconsin, USA). The setup employed collimated light, achieved by removing the bottom condenser, and a liquid-nitrogen-cooled mercury-cadmium-telluride (MCT) detector for spectral measurements. All acquired transmittance spectra of the MHA gradient metasurfaces were normalized to the transmittance of standard CaF$_2$ windows.

**Data processing**
The measured transmittance spectrum in Figure 2b (experiment) was normalized to its maximum for visualization purposes. All cavity losses and Q-factors were determined by fitting the raw transmittance data shown in Figure 2b to the following Fano expression.[38]:

$$T = \left| ie^{i\phi}t_0 + \frac{\Gamma_R}{\Gamma_R + \Gamma_{NR} + i(\lambda - \lambda_{res})} \right|^2$$

Where $ie^{i\phi}t_0$ describes the background and the shape of the resonance, with the asymmetry factor $\alpha$ being defined as $t_0$. The Q-factor can then be calculated as:

$$Q = \frac{\lambda_{res}}{2(\Gamma_R + \Gamma_{NR})}$$

Where $\lambda_{res}$, $\Gamma_R$, and $\Gamma_{NR}$ are the resonance wavelength, radiative, and absorptive loss rates, respectively.

## Data availability
The data that support the findings of this study are available from the corresponding author upon reasonable request.

## Competing Interest
The authors declare no competing interest.

## Acknowledgement

The authors thank Professor Eduardo R. Arvelo, affiliated with the Electrical & Computer Engineering Department at UW-Madison, for his assistance with 3D modeling and rendering. The authors also thank PhD student Justin Edwards for facilitating a macro lens to take close-up photographs of the MHA gradient metasurface. The authors gratefully acknowledge use of facilities and instrumentation in the UW-Madison Wisconsin Center for Nanoscale Technology. The Center (wcnt.wisc.edu) is partially supported by the Wisconsin Materials Research Science and Engineering Center (NSF DMR-2309000) and the University of Wisconsin-Madison. The authors thank Professor Mikhail Kats for providing access to the FTIR. The MHA $Si_3N_4$ membranes were gold-coated at the Center for Nanoscale Materials at the Argonne National Laboratory, a U.S. Department of Energy Office of Science User Facility, was supported by the U.S. DOE, Office of Basic Energy Sciences, under Contract No. DE-AC02-06CH11357. F.Y. acknowledges financial support from the U.S. National Science Foundation (grant no. 2401616) and the U.S. National Institutes of Health (grant no. R21EB034411).